\documentclass[doublecol]{epl2}
\usepackage{subfigure}
\usepackage{epic}\usepackage{eepic}
\usepackage{bm}
\usepackage[dvips]{epsfig}
\usepackage{graphicx} \usepackage{amsmath} \usepackage{amssymb}
\usepackage[colorlinks=true,urlcolor=magenta,citecolor=blue,linkcolor=blue]{hyperref}
\newcommand{\comment}[1]{}

\renewcommand{\d}{{\rm d}}

\newcommand{\Z}{{\cal Z}}
\newcommand{\F}{{\cal F}}
\newcommand{\p}{\bar{p}}
\renewcommand{\vec}[1]{\boldsymbol{#1}}
\renewcommand{\a}{\vec{a}}
\newcommand{\w}{\vec{w}}

\title{Free energy for non-equilibrium quasi-stationary states}

\author{A.E. Allahverdyan and N.H. Martirosyan} \institute{ Yerevan
  Physics Institute, Alikhanian Brothers Street 2, Yerevan 375036,
  Armenia} \shortauthor{Allahverdyan and Martirosyan}
 
\abstract{  We study a class of non-equilibrium quasi-stationary states for a Markov system
interacting with two different thermal baths. We show that the work done under a slow, external
change of parameters admits a potential, i.e., the free energy. Three conditions are needed for the
existence of free energy in this non-equilibrium system: time-scale separation between variables
of the system, partial controllability (external fields couple only with the slow variable), and an
effective detailed balance. These conditions are facilitated in the continuous limit for the slow
variable. In contrast to its equilibrium counterpart, the non-equilibrium free energy can increase
with temperature. One example of this is that entropy reduction by means of external fields
(cooling) can be easier (in the sense of the work cost) if it starts from a higher temperature.
}


\begin{document}

\maketitle

\textit{\textbf{ Introduction.}}  
One reason for the effectiveness of
thermodynamics is that its concepts and ideas apply beyond
the domain of equilibrium states, e.g., in real life
thermodynamics is applied to systems having different
temperatures, even though such states do not belong to
equilibrium. There are specific mechanisms for this applicability,
e.g., thermodynamics applies to systems that
are perturbatively close to equilibrium \cite{zubarev,strat,meixner,balian}. Another
general mechanism is the time-scale separation. In fact,
this mechanism is inherent in the structure of thermodynamics,
which is built up via the notion of a quasi-static
process \cite{meixner}. It is also important in statistical physics of
open systems, where even low-dimensional systems can
play the role of a thermal bath, provided that they are
fast \cite{jar,baba}. A related point is seen in stochastic thermodynamics,
where the notion of white (i.e., fast) noise is
relevant \cite{sekimoto}.

There is already a body of work concerning thermodynamic
aspects of systems with time-scale separation
\cite{cug_kurch_pel,casas,cug,bertin,theo,bo,espo,pugo,gomez,tomo,ziener,
  landauer,an,sg1,sg2,sg3,karen,figur,ilg}. In particular, much attention was devoted
to the effective temperature of glassy systems, where
the time-scale separation emerges from the many-body
physics \cite{cug_kurch_pel,theo,cug,bertin}. Recent works studied time-scale separation
in stochastic thermodynamics focusing on dissipative
features such as entropy production \cite{bo,espo,pugo,gomez,tomo,ziener}.

Time-scale separation is a much wider notion, and it is frequent also
in biology and society, where it allows to reduce the complexity of
emergent structures \cite{roj,gun}.  Indeed, components of such
systems enjoy a certain autonomy (though in different ways): fast
variables ``live'' under fixed values of the slow ones, while the slow
variables ``see'' stationary distributions of the fast ones.

% The majority of of applications of thermodynamic concepts (such as
% entropy or temperature) in non-equilibrium situation are such that
% these concepts behave effectively similar to their equilibrium
% counterparts. For entropy and/or temperature the equlibrium analogy is
% a necessary part of their definition, since these are emergent
% (non-mechanic) concepts \cite{casas,cug}.

We aim to look at a class of stationary, non-equilibrium states for a
Markov stochastic system that is out of equilibrium due to interaction
with two different thermal baths; see \cite{shaw} for an introduction
to such systems. We impose three conditions.

-- Time-scale separation: the system under consideration consists of
two variables, fast and slow. 

% This assumption comes out naturally if the to-be-slow variable is
% quasi-continuous, while the other one is discrete.

-- Partial controllability: external fields act on the slow variable
only. This assumption can be validated from operational reasons: it is
normally difficult to have a precise control on variables that move
fast.
\begin{figure*}[htbp]
\centering
\vspace{0.1cm}
%\subfigure[]{
    \includegraphics[width=8.5cm]{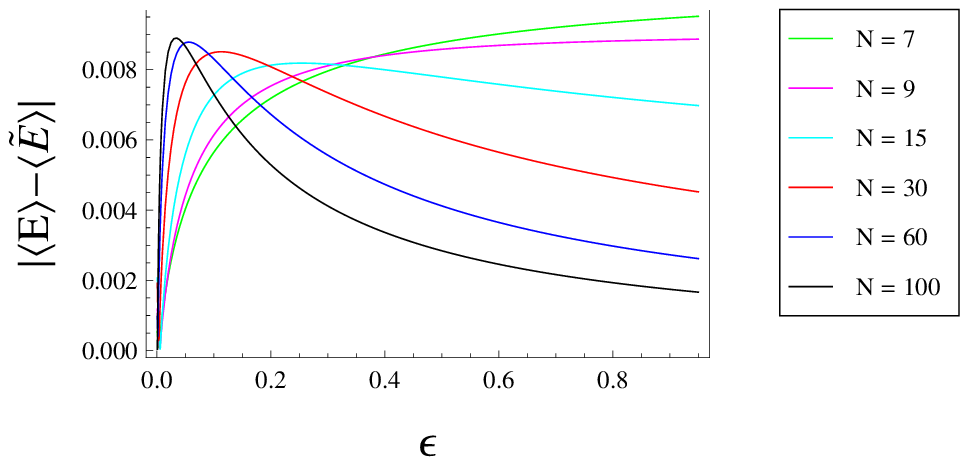} 
%\label{f11}
%}
%\subfigure[]{
    \includegraphics[width=8.5cm]{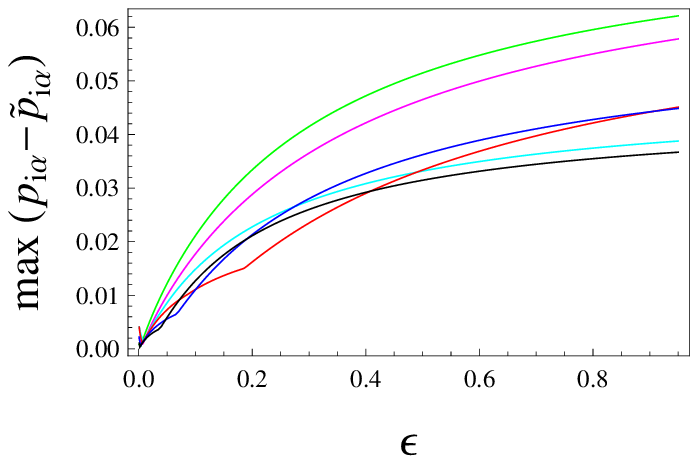}
%\label{f12}
%}
    \caption{TТime-scale separation approach versus numerically exact
      features of the stationary distribution given by (\ref{eq:1}).
      In (\ref{eq:2}, \ref{eq:3}) we took $\beta=1$, $\beta_s=10$,
      $E_{i\alpha}={\alpha i^2}/{N}$ and $n=2$. Here $N$ ($n$) is the
      number of different states for the slow (fast) variable. For
      transition rates in (\ref{eq:2}, \ref{eq:3}) we choose the
      Kawasaki rates: $\rho_{ij|\alpha}=e^{\frac{\beta}{2}( E_{j\alpha
        }-E_{i\alpha }) }$ and
      $\omega_{\alpha\gamma|i}=e^{\frac{\beta_s}{2}( E_{i\gamma
        }-E_{i\alpha
        })  }  $. 
      {\bf Left panel}: $|\langle E\rangle- \langle\widetilde{E}\rangle|$
      vs. $\epsilon$, where $\langle E\rangle=\sum_{i\alpha}
      p_{i\alpha} E_{i\alpha}$ is the exact average energy in the
      stationary state, $\langle\widetilde{E}\rangle$ is the average
      energy calculated via the time-scale separation approach, and
      $\epsilon$ controls the slow-fast limit in (\ref{eq:1}). The
      value of $\langle E\rangle$ (not shown on figure) is some
      100--150 times larger than $|\langle
      E\rangle- \langle\widetilde{E}\rangle|$.   
      {\bf Right panel}: ${\rm max}(p_{i\alpha}-\widetilde{p}_{i\alpha})$
      vs. $\epsilon$, where $p_{i\alpha}$ are the exact stationary
      probabilities, while $\widetilde{p}_{i\alpha}$ are calculated
      via the time-scale separation approach. We found that ${\rm
        max}(p_{i\alpha}-\widetilde{p}_{i\alpha})={\rm
        max}|p_{i\alpha}-\widetilde{p}_{i\alpha}|$. }
\end{figure*}
Both conditions need not hold for the equilibrium situation (equal
temperatures of the thermal baths), where thermodynamics applies under
any type of controllability and the ratio of characteristic
times. 

-- Transition rates of the slow variable hold certain
constraints|they are given by activation rates, or the slow variable
lives in a tree-like structure|that amount to an effective detailed
balance that holds after averaging (tracing out) over the fast
variable.

Under these conditions the slow (isothermal) work done on this system
admits a potential, i.e., there exists the free energy.  This
definition is unambiguous, because the work is defined for an
arbitrary non-equilibrium state \cite{balian}; see the discussion
after (\ref{eq:24}) for details of this point. Hence the free
energy need not have further equilibrium features. Indeed, we show
that its behavior with respect to temperature changes can be different
from the features of the equilibrium free energy.

\comment{
\begin{figure}[htbp]
%\vspace{-1cm}
\includegraphics[width=9cm]{two_temp_0.eps}
\vspace{-2.5cm}
\caption{Schematic representation of the model. Greek and Latin
  indices denote slow and fast variables, respectively. Bold (normal)
  arrows indicate on fast (slow) transitions. }
\end{figure}}

\textit{\textbf{ Two-temperature Markov dynamics.}}  Consider the
following master equation for two discrete variables $i=1,...,n$ and
$\alpha=1,...,N$ (see, e.g., \cite{strat,sekimoto}):
\begin{eqnarray}
  \dot{p}_{i\alpha}={\sum}_j [\rho_{ij|\alpha}\,p_{j\alpha}   
  -\rho_{ji|\alpha}\,p_{i\alpha}
  ]\nonumber\\
  +\epsilon
  {\sum}_{\gamma}[\omega_{\alpha\gamma|i}\,
  p_{i\gamma}
  -\omega_{\gamma\alpha|i}\, p_{i\alpha}  ], 
  \label{eq:1}
\end{eqnarray}
where $p_{i\alpha}$ is the joint probability of $i$ and $\alpha$,
$\rho_{ij|\alpha}$ and $\omega_{\alpha\gamma|i}$ are the transition
probabilities for $i\leftarrow j$ (for a fixed $\alpha$) and
$\alpha\leftarrow \gamma$ (for a fixed $i$), respectively. In
(\ref{eq:1}), $\epsilon$ is a small parameter that makes $\{\alpha\}$
slower than $\{i\}$. We assume that all sums over Latin (Greek)
indices run from $1$ to $n$ (from $1$ to $N$).

The transitions are controlled by different thermal baths at
temperatures $T=1/\beta>0$ and $T_s=1/\beta_s>0$ respectively. Hence
the transition probabilities $\rho_{ij|\alpha}$ and
$\omega_{\alpha\gamma|i}$ hold the detailed balance conditions (see
e.g., \cite{strat,sekimoto}):
\begin{eqnarray}
  \label{eq:2}
  \rho_{ij|\alpha}~e^{-\beta E_{j\alpha
    }}=\rho_{ji|\alpha} ~e^{-\beta E_{i\alpha }}, \\
  \label{eq:3}
  \omega_{\alpha\gamma|i}~e^{-\beta_s E_{i\gamma
    }}=\omega_{\gamma\alpha|i} ~e^{-\beta_s E_{i\alpha }}.
\end{eqnarray}
Without loss of generality we parametrize (\ref{eq:3}) as
\begin{eqnarray}
  \label{eq:10}
  \omega_{\alpha \, \gamma|i}=e^{B_{\alpha \,
      \gamma|i}+\frac{\beta_s}{2}
    (E_{i\, \gamma}-E_{i\, \alpha})}, \qquad   B_{\alpha\,
      \gamma|i}=B_{\gamma  \,
     \alpha|i},
\end{eqnarray}
where $B_{\alpha\,
      \gamma|i}$ accounts for the symmetric part of 
    $\alpha\leftrightarrow \gamma$.

\textit{Time-scale separation} holds when $\epsilon$ in
(\ref{eq:1}) is sufficiently small; see fig.~1. This is a reliable approximation,
since its predictions are close to the exact stationary
probability even for moderately small $\epsilon$; see the left panel of fig.~1

We now work out the stationary state ($\dot{p}_{i\alpha}=0$) of
(\ref{eq:1}) for a small $\epsilon$ following the standard
perturbation theory approach; \cite{pavlo} for a rigorous presentation
and \cite{kampen} for a physical discussion. One puts into 
(\ref{eq:1})
\begin{eqnarray}
  \label{eq:13}
  p_{i\alpha}=B_{i\alpha}^{[0]}+{\sum}_{a\geq 1}\, \epsilon^a B_{i\alpha}^{[a]},
\end{eqnarray}
and obtains for successive terms which are of order ${\cal O}(1)$ and
${\cal O}(\epsilon^a)$, respectively:
\begin{align}
  \label{eq:27}
&  {\sum}_j [\rho_{ij|\alpha}\,B^{[0]}_{j\alpha}
  -\rho_{ji|\alpha}\,B^{[0]}_{i\alpha}]=0, \\
&  {\sum}_j [\rho_{ij|\alpha}\,B^{[a]}_{j\alpha}
  -\rho_{ji|\alpha}\,B^{[a]}_{i\alpha}]\nonumber\\
&+{\sum}_{\gamma}
  [\omega_{\alpha\gamma|i}\, B^{[a-1]}_{i\alpha}
  -\omega_{\gamma\alpha|i} \, B^{[a-1]}_{i\alpha} ]=0,\quad a\geq 1.
\label{aa}
\end{align}
Note that there are solvability conditions found from summing
(\ref{aa}) over $i$:
\begin{eqnarray}
  \label{eq:31}
  {\sum}_{i\gamma}
  [\omega_{\alpha\gamma|i}\, B^{[a-1]}_{j\alpha}
  -\omega_{\gamma\alpha|i} \, B^{[a-1]}_{j\alpha} ]= 0,\quad a\geq 1.
\end{eqnarray}
Now (\ref{eq:27}) and (\ref{eq:31}) with $a=1$ are solved as 
\begin{eqnarray}
  \label{eq:32}
  B_{i\alpha}^{[0]}=\bar{p}_{i|\alpha}\bar{p}_{\alpha},
\end{eqnarray}
where we used (\ref{eq:2}), the stationary conditional probability
$\bar{p}_{i|\alpha}$ of the fast variable has the equilibrium form
\begin{eqnarray}
  \label{eq:7}
  \bar{p}_{i|\alpha}=e^{-\beta E_{i\alpha}}\left/Z_\alpha[\beta]\right., \qquad 
Z_{\alpha}[\beta]={\sum}_k\, e^{-\beta E_{k\alpha}},
\end{eqnarray}
and where the probability $\bar{p}_{\alpha}$ of the slow variable is
found from (\ref{eq:31}) with $a=1$
\begin{eqnarray}
\label{eq:x5}
%\dot{p}_{\alpha}=\epsilon
{\sum}_{\gamma} [\bar\Omega_{\alpha\gamma}\,
p_{\gamma} -\bar\Omega_{\gamma\alpha}\, p_{\alpha}  ]=0, ~~
\bar\Omega_{\alpha\gamma}\equiv{\sum}_i \,\omega_{\alpha\gamma|i}\,
\bar p_{i|\gamma}. 
\end{eqnarray}
The meaning of (\ref{eq:32}) is that the fast variable relaxes to the
conditional equilibrium, which determines via (\ref{eq:x5}) and the
effective rates $\bar\Omega_{\alpha\gamma}$ the probability of the
slow variable.

Due to $\sum_{i\alpha}p_{i\alpha}=\sum_{i\alpha}\bar p_{i|\alpha} \bar
p_{\alpha}=1$, (\ref{eq:32}) implies normalization condition:
$\sum_{i\alpha}B_{i\alpha}^{[a]}=0$ for $a\geq 1$. Thus
$B_{i\alpha}^{[1]}$ is found from this condition, (\ref{aa}) with
$a=1$, and (\ref{eq:31}) with $a=2$. Expectedly, for a small
$\epsilon$ we get $B_{i\alpha}^{[1]}={\cal O}(\epsilon)$. Now the
right panel of fig.~1 confirms this fact with numeric results, but it also shows
that the difference between (\ref{eq:32}) and the true joint
probability $p_{i \alpha}$ is a sublinear function of $\epsilon$ for a larger values of
$\epsilon$. This is favorable for the approximation
\begin{eqnarray}
  \label{eq:333}
  p_{i\alpha}=  \bar{p}_{i|\alpha}\bar{p}_{\alpha},
\end{eqnarray}
that we adopt from now on for the stationary probability $p_{i\alpha}$
of (\ref{eq:1}).

\textit{External fields} are introduced via time-dependent
parameters $\a(t)=(a_1(t),...,a_A(t))$ in the energy $E_{i\alpha}(\a)$
of the system.  We assume that
\begin{eqnarray}
  \label{eq:11}
  E_{i\alpha}(\a)=E_{\alpha}(\a)+\hat E_{i\alpha},
\end{eqnarray}
where $\hat E_{i\alpha}$ does not depend on $\a$. Eq.~(\ref{eq:11})
means that external fields couple only with the slow variable,
e.g., because it is difficult to control fast objects. The
stationary probabilities (\ref{eq:333}) depend on parameters $\a$
through (\ref{eq:11}).

Now $\a(t)$ is slow as compared to the relaxation of both fast and
slow variables.  The temperatures $T$ and $T_s$ are constant. Hence
the (quasi-stationary) probabilities of the system are found from
(\ref{eq:333}), where $\a$ is replaced by $\a(t)$. The differential
thermodynamic work $\w$ is \cite{strat} [see (\ref{eq:11})]
\begin{eqnarray}
  \label{eq:14}
 \w ={\sum}_{i\alpha}\, \bar{p}_{i\alpha} \,\partial_{\a} E_{i\alpha}
  ={\sum}_\alpha\, \bar{p}_\alpha \,\partial_{\a}
  E_\alpha,
\end{eqnarray}
where $\partial_{\a}$ is the gradient in the $\a$-space. Note that
relations between thermodynamic and mechanic work have to be specified
in the context of concrete applications \cite{rubi}. This has to do
with the following freedom in the definition (\ref{eq:14}):
$\partial_{\a} E_\alpha$ (and hence $\w$) will change upon adding to
$E_\alpha$ a factor $\varphi (\a)$ that does not depend on $\alpha$,
but depends on $\a$ (this is akin to the gauge-freedom of the
potential energy in mechanics). Hence the energies $E_\alpha$ need to
be specified, before defining the work $\w$ \cite{rubi}.

Each component of $\w=(w_1,...,w_A)$ can have a separate physical
meaning, since components of $\a$ may be driven by different sources.

The integral work is a line integral in the $\a$-space:
\begin{eqnarray}
  \label{eq:24}
  W=\int_{t_{\rm in}}^{t_{\rm f} } \d t\,\frac{\d \a}{\d t}\,\,\w(t) =
\int_{\a_{\rm in}}^{\a_{\rm f} } \d \a\,\,\w ,
\end{eqnarray}
where $\a_{\rm in}=\a(t_{\rm in})$ and $\a_{\rm f}=\a(t_{\rm f})$ are
the initial and final values of $\a(t)$ reached at times $t_{\rm in}$
and $t_{\rm f}$, respectively. We stress that (\ref{eq:24}) refers to
slow changes of parameters $\a(t)$, but if we change
$\bar{p}_{i\alpha}$ in (\ref{eq:14}) to the time-dependent probability
$p_{i\alpha}$ found from (\ref{eq:1}), then the same expression for
work applies for arbitrary processes \cite{balian}.

The work admits a potential (i.e., free energy $\F$) if
\begin{eqnarray}
  \label{canton}
\w=\partial_{\a} \F, \qquad W= \F(\a_{\rm f})-\F(\a_{\rm in}).
\end{eqnarray}
If (\ref{canton}) does not hold, the work extraction (i.e., $W<0$) by
means of a slow, cyclic ($\a_{\rm in}=\a_{\rm f}$) variation of $\a$
is possible. Indeed, if $W\not=0$, then $W$ changes its sign when the
cycle is passed in the opposite direction.  The extracted work is
determined by the closed-contour integral.

Below we give pertinent examples of $W\not=0$ for cyclic processes;
see (\ref{eq:19}).  Now we turn to studying cases, in
which (\ref{canton}) does hold despite the fact that $T_s\not= T$.

\textit{\textbf{ Non-equilibrium free energy for the activation
    rate.}}  An example of slow dynamics is given by the activation
energy rate in (\ref{eq:10}) \cite{sekimoto}
\begin{eqnarray}
  \label{eq:16}
\omega_{\alpha \, \gamma|i}=e^{\beta_sE_{i\gamma}}, ~~
  B_{\alpha \,
    \gamma|i}=  {\beta_s}(E_{i\gamma}+E_{i\, \alpha})/2.
\end{eqnarray}
One interpretation of (\ref{eq:16}) is that there is a barrier with
energy $E^*$ that is larger than all other energies. Choosing 
$E^*=0$, we see that $\omega_{\alpha \, \gamma|i}$ in (\ref{eq:16})
assumes the standard Arrhenius form. 

Now $\bar\Omega_{\alpha\gamma}$ in (\ref{eq:x5}) depends only on
$\gamma$: $\bar\Omega_{\alpha\gamma}=\bar\Omega_\gamma$. The
probability $\bar{p}_\alpha$ in (\ref{eq:x5}) is found via the
detailed balance condition $\bar\Omega_{\alpha\gamma} \bar p_{\gamma}
=\bar\Omega_{\gamma\alpha}\bar p_{\alpha}$. This leads to $\bar
p_{\gamma} \propto 1/\bar\Omega_{\gamma}$. Thus we get from
(\ref{eq:7}, \ref{eq:x5}, \ref{eq:11}, \ref{eq:16})
\begin{eqnarray}
  \label{eq:73}
\bar p_{\alpha}={\mu_\alpha e^{-\beta_s E_\alpha}}\left/{{\sum}_\gamma\,
  \mu_\gamma e^{-\beta_s E_\gamma}}\right.,
\end{eqnarray}
where $\mu_\alpha$ corresponds to the weight of the energy $E_\alpha$
and expresses via the statistical sum of the fast variable:
\begin{eqnarray}
\label{au}
\mu_{\alpha}&=&\hat Z_\alpha[\beta]\, /\,
  \hat Z_\alpha[\beta-\beta_s]  , \\
\label{bau}
\hat Z_\alpha[\beta]&\equiv&{\sum}_k e^{-\beta \hat E_{k\alpha}}.
\end{eqnarray}
Note that $\mu_\alpha$ does not depend on $\a$ due to (\ref{bau}) and
(\ref{eq:11}).  Eqs.~(\ref{eq:11}, \ref{canton}, \ref{eq:73}) imply
the existence of a free energy:
\begin{eqnarray}
  \label{eq:42}
  \F=-T_s\ln\left[{\sum}_\alpha\,\, \mu_\alpha \,e^{-\beta_s E_\alpha}
  \right].
\end{eqnarray}
Recall that $\F$ is defined up to a constant, 
\begin{eqnarray}
  \label{arbitrary}
\F\to\F+{\cal C},   
\end{eqnarray}
that can depend on anything besides $\a$
\cite{rubi}. Eq.~(\ref{eq:42}) is obtained under a specific choice of
${\cal C}$ that proves useful below when calculating derivatives of
(\ref{eq:42}).

% Note that for $\beta<\beta_s$, the statistical sum in
% (\ref{au}) is calculated at a negative temperature.

\textit{\textbf{ Temperature-dependence of $\F$and entropy reduction
    (cooling) via external fields.}} The free energy $F_{\rm
  eq}(\a)=-T\ln\sum_k e^{-\beta E_k}$ of an equilibrium system is a
decreasing function of the temperature:
\begin{eqnarray}
  \label{star}
{\partial_T F}_{\rm eq}(\a)=-S_{\rm eq}(\a)\leq 0,
\end{eqnarray}
where $S_{\rm eq}(\a)$ is the entropy. Since $F_{\rm eq}(\a)$ is
defined under a specific choice of an additive {\it and}
temperature-dependent constant (cf.~(\ref{arbitrary})), (\ref{star})
changes for a different choice of the constant. Hence (\ref{star}) cannot
be interpreted directly. But it can be related to the work cost of a
cooling process via externally driven parameters $\a(t)$ \cite{jan}.

Normally, cooling means temperature reduction of a macroscopic system
that has a single and well-defined temperature. But the needs of
NMR-physics \cite{nmr_cooling}, atomic and molecular physics
\cite{ket,kosloff}, quantum computation \cite{fernan_cooling} {\it
  etc} led to generalizing this definition \cite{jan,dvira,popo}.  In
these fields one needs to reduce the entropy of a system that is
coupled to a fixed-temperature thermal bath. Reducing the bath
temperature is not feasible. But it is feasible to reduce the entropy
via external fields.  For instance, in NMR spin systems, entropy
decrease for a spin means its polarization increase, which is
necessary for the NNR spectroscopy \cite{nmr_cooling}.

Thus, cooling amounts to an isothermal process, where $\a(t)$ slowly
changes from $\a_{\rm in}$ to $\a_{\rm f}$ and achieves a lower final
entropy. For the equilibrium situation this means $S_{\rm eq}(\a_{\rm
  f})< S_{\rm eq}(\a_{\rm in})$ (dynamic aspects of this problem are
analyzed in \cite{jan}). Now
\begin{eqnarray}
  \label{eq:15}
  \partial_T [F_{\rm eq}(\a_{\rm f})-F_{\rm eq}(\a_{\rm in})]=S_{\rm eq}(\a_{\rm
    in})-S_{\rm eq}(\a_{\rm f})\geq 0,  
\end{eqnarray}
compares two setups at different temperatures, but the same values
$\a_{\rm in}\to\a_{\rm f}$ of the external fields. Eq.~(\ref{eq:15})
means that the work cost $F_{\rm eq}(\a_{\rm f})-F_{\rm eq}(\a_{\rm
  in})$ of cooling increases with the temperature $T$, i.e., cooling
from a higher temperature is harder, as expected.

Turning to the non-equilibrium free energy $\F$, we characterize its
temperature dependence via $\partial_{T_s} \F|_{T}$ and $\partial_{T}
\F|_{T_s}$, since $T$ and $T_s$ are independent parameters. We deduce
for the activation energy rate (\ref{eq:16}, \ref{au}):
\begin{align}
\label{eq:77}
&\partial_{T_s} \F|_{T}=-S_s- {\sum}_\alpha\, \p_\alpha \left[
  \beta_s\hat E_\alpha(\beta-\beta_s)+\ln\mu_\alpha \right]\\
&=-S_s-{\sum}_\alpha\, \p_\alpha\int_0^{\beta_s} \d y\, [\,\hat
    E_\alpha(\beta-\beta_s)-
    \hat E_\alpha(\beta-y) \,],\nonumber\\
  \label{eq:78}
&  \partial_{T} \F|_{T_s}=\frac{\beta^2}{\beta_s}
{\sum}_\alpha \p_\alpha
  \left[\,\hat
    E_\alpha(\beta-\beta_s)-
    \hat E_\alpha(\beta)
  \right],
\end{align}
where 
\begin{eqnarray}
  S_s=-\sum_\alpha \bar p_\alpha\ln \bar p_\alpha
\end{eqnarray}
 is the entropy
of the slow variable (cf.~(\ref{star})), and 
\begin{eqnarray}
  \hat
E_\alpha=\frac{1}{\hat Z(\beta)}\sum_k \hat E_{k\alpha}e^{-\beta \hat
  E_{k\alpha}}
\end{eqnarray}
 is the conditionally averaged energy of the fast
variable; cf.~(\ref{eq:7}). $\hat E_\alpha (\beta)$ monotonously
decays from ${\rm max}_k[E_{k\alpha}]$ to ${\rm min}_k[E_{k\alpha}]$
when $\beta$ goes from $-\infty$ to $\infty$. Hence we get in
(\ref{eq:77}, \ref{eq:78}):
\begin{eqnarray}
  \label{eq:25}
  \partial_{T_s}
  \F|_{T}\leq 0, \qquad \partial_{T} \F|_{T_s}\geq 0.   
\end{eqnarray}

Now we explore implications of $\partial_{T} \F|_{T_s}\geq 0$ for a
cooling process. Denote by
\begin{eqnarray}
  \label{eq:28}
  \hat S_{\alpha}(\beta)= -
  {\sum}_i \bar p_{i|\alpha} \ln\bar p_{i|\alpha}=\beta\hat
  E_\alpha+\ln \hat Z_\alpha,
\end{eqnarray}
the entropy of the fast variable conditioned by a fixed value $\alpha$
of the slow variable; cf.~(\ref{eq:7}). Recall that $S_s+\sum_\alpha
\bar p_\alpha\hat S_{\alpha}(\beta)$ amounts to the full entropy $-
\sum_{i\alpha} \bar p_{i\alpha} \ln\bar p_{i\alpha}$.

We denote: $\hat S_1={\rm min}_\alpha[\hat S_\alpha]$. Then we can
define a cooling process, where all energies $E_{\alpha\not =1}$ in
(\ref{eq:11}) slowly increase leading to $\p_1\to 1$; see
(\ref{eq:73}).  Though the process is realized by external fields
holding the partial controllability restriction (\ref{eq:11}), we
still get a cooling of the {\it whole} (slow plus fast)
system. Indeed, not only the entropy $S_s$ of the slow variable
decreases to zero, but also the conditional entropy of the fast
variable decreases from its initial value $\sum_\alpha \bar
p_\alpha\hat S_{\alpha}(\beta)$ to a smaller value $\hat
S_{1}(\beta)$.

We note that the considered cooling process can also decrease the
internal energy of the system. Recall that the internal energy is
defined as [cf.~(\ref{eq:333}), (\ref{eq:11})]
\begin{eqnarray}
  \label{eq:4}
{\sum}_{i\alpha}E_{i\alpha}\bar p_{i|\alpha}\bar p_{\alpha}=
{\sum}_{\alpha}E_{\alpha}\bar p_{\alpha}+{\sum}_{\alpha}\hat
E_{\alpha}\bar p_{\alpha}, 
\end{eqnarray}
where $\hat E_\alpha$ is defined after (\ref{eq:78}). Since $\hat
S_\alpha$ and $\hat E_\alpha$ are equilibrium quantities, $\hat
S_\alpha$ is an increasing function of $\hat E_\alpha$. Hence it is
possible to choose $\hat E_1={\rm min}_\alpha[\hat E_\alpha]$ in
addition to $\hat S_1={\rm min}_\alpha[\hat S_\alpha]$. We can also
choose $E_1={\rm min}_\alpha[\hat E_\alpha]$. Then $\p_1\to 1$ means
that the internal energy (\ref{eq:4}) decreases during the cooling.

The cooling process incurs a work cost [cf.~(\ref{eq:42})]
\begin{eqnarray}
  \label{eq:30}
\Delta \F=-T_s\ln[\mu_1e^{-\beta E_1}]-\F\geq 0,
\end{eqnarray}
where $\F$ is the initial free energy. Now $\Delta\F>0$ means that the
work is taken from the external source, hence this is indeed a work
cost.

The change of $\Delta \F$ with the temperature $T$ of the fast
variable reads from (\ref{eq:78}):
\begin{align}
\partial_{T}\Delta \F|_{T_s}=-\frac{\beta^2}{\beta_s}
{\sum}_\alpha \p_\alpha
  \left[\,\hat
    E_\alpha(\beta-\beta_s)-
    \hat E_\alpha(\beta) \right.\nonumber \\ \left. 
-\hat E_1(\beta-\beta_s)+\hat E_1(\beta)
  \right].
\label{wot}
\end{align}
Let us now indicate several scenarios for $\partial_{T}\Delta
\F|_{T_s}\leq 0$ in (\ref{wot}), and show that they are consistent
with condition $\hat S_1={\rm min}_\alpha[\hat S_\alpha]$ that defines
the cooling process. For example, in (\ref{wot}) one can take $\hat
E_1(\beta-\beta_s)\simeq\hat E_1(\beta)$, but $\hat
E_{\alpha\not=1}(\beta-\beta_s)\not\simeq\hat
E_{\alpha\not=1}(\beta)$. Another example is to make $\beta_s$
small. Then $E_\alpha(\beta-\beta_s)- \hat E_\alpha(\beta)\propto \hat
C_\alpha$ amounts to the heat-capacity $\hat C_\alpha\geq 0$, and then
$\partial_{T}\Delta \F|_{T_s}<0$ can be implied by $\hat C_{\alpha\not
  =1}>\hat C_1$, which is consistent with $\hat S_1={\rm
  min}_\alpha[\hat S_\alpha]$.

We conclude from $\partial_{T}\Delta \F|_{T_s}\leq 0$ that it can be
easier (in terms of the work cost) to cool from higher temperatures
than from the lower ones. Note that the result survives also in the
near-equilibrium limit $\beta\simeq\beta_{s}$. With the same logics,
one shows from (\ref{eq:77}) that when increasing the temperature
$T_s$ of the slow variable, one normally has $\partial_{T_s}\Delta
\F|_{T}\geq 0$ [cf.~(\ref{eq:15})], and we revert to equilibrium with
$[\partial_{T_s}\Delta \F|_{T}+\partial_{T}\Delta
\F|_{T_s}]_{T=T_s}\geq 0$.

% According to (\ref{eq:88}), $\F$ increases with the temperature $T$ of
% the fast degree of freedom. This is just opposite to what is expected
% from the equilibrium analogy. $\F$ can also increase as a function of
% $T_s$. The takes place just in the quasi-equilibrium situation
% $\beta_s\lesssim\beta$; see (\ref{eq:77}).

\comment{  $\hat S_\alpha (\beta)$, $\hat E_\alpha (\beta)$ and $\hat
  F_\alpha (\beta)=\hat E_\alpha (\beta)-T\hat S_\alpha
  (\beta)=-T\ln\hat Z_\alpha$ are, respectively, equilibrium entropy,
  average energy and free energy of the fast system with energies
  $\{\hat E_{k\alpha}\}_k$ and temperature $T=1/\beta$, e.g., $\hat
  E_\alpha=\frac{1}{\hat Z(\beta)}\sum_k \hat E_{k\alpha}e^{-\beta
    \hat E_{k\alpha}}$. We recall features of $\hat E_\alpha (\beta)$,
  $\hat S_\alpha (\beta)$ and $\hat F_\alpha (\beta)$ as functions of
  a real (positive or negative) $\beta$. $\hat E_\alpha (\beta)$
  monotonously decreases with $\beta$ from ${\rm max}_k[E_{k\alpha}]$
  at $\beta=-\infty$ to ${\rm min}_k[E_{k\alpha}]$ at $\beta=\infty$.
  $\hat S_\alpha (\beta)\geq 0$ attains its single maximum at
  $\beta=0$. It can nullify only for $\beta\to\pm \infty$. $\hat
  F_\alpha(\beta)$ consists of two branches for $\beta>0$ and
  $\beta<0$. Within each branch $\partial_\beta F_\alpha(\beta)>0$,
  and $\hat F_\alpha(-|\beta_1|)>\hat F_\alpha(|\beta_2|)$ for any
  $\beta_1$ and $\beta_2$.  }

% $\hat F_\alpha(\beta\to 0)\to -{\rm sign}[\beta]\times\infty$, while
% $\hat F_\alpha(\beta\to \infty)\to {\rm min}_k[E_{k\alpha}]$ and $\hat
% F_\alpha(\beta\to -\infty)\to {\rm max}_k[E_{k\alpha}]$.

\comment{On very long times any isolated two-temperature system is
quasi-stationary, since temperatures $T$ and $T_s$ tend to equalize
when the overall system goes to global equilibrium. Eqs.~(\ref{eq:88})
show that $\F$ can increase during this process.}

\begin{figure*}[htbp]
\centering
    \includegraphics[width=15cm, height=4.0cm]{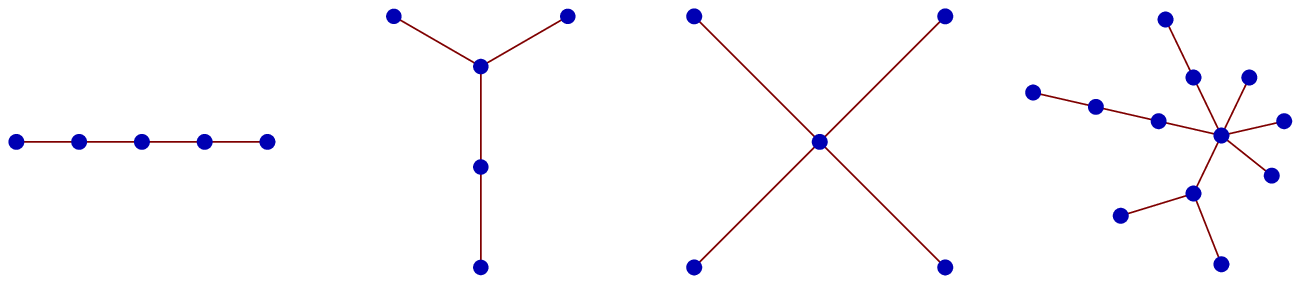}
\vspace{0.3cm}
\caption{Four examples of tree-like structures. Bold points denote
  states, and lines between them indicate on inter-state
  transitions. }
\end{figure*}

\textit{\textbf{ Tree-like topology of the slow variable.}}  Free
energy (\ref{eq:42}) exists for general rates (\ref{eq:10}), if the
topology of connections between the states $\alpha, \gamma,...$ in
(\ref{eq:x5}) is that of a tree (a network without loops or closed
cycles): for a fixed $i$, $\omega_{\alpha \gamma|i}\not=0$ (and hence
$\bar\Omega_{\alpha\gamma}\not=0$) only along branches of a tree; see
fig.~2 for examples. Let node $\sigma_0$ be the root of the tree. This
root can be chosen arbitrarily, the freedom of choosing it will
connect to an arbitrary constant (\ref{arbitrary}) in the free energy.
Each node $\sigma$ of the tree is related to $\sigma_0$ via a unique
path $\sigma\sigma'...\sigma'' \sigma_0$. The stationary probability
$\bar p_\sigma$ of $\sigma$ is deduced from (\ref{eq:x5}), it is made
of the transition probabilities along this unique path:
\begin{eqnarray}
  \label{eq:63}
  \bar p_\sigma=\frac{\bar\Omega_{\sigma\sigma'}
    \bar\Omega_{\sigma' ...} \bar\Omega_{...\sigma''} \bar\Omega_{\sigma''
      \sigma_0}}
{\bar\Omega_{\sigma'\sigma}
    \bar\Omega_{...\sigma' } \bar\Omega_{\sigma'' ...} \bar\Omega_{\sigma_0
      \sigma'' }}
\,\bar p_1,
\end{eqnarray}
where $\bar p_1$ is gotten from normalization. We add to
(\ref{eq:11}):
\begin{eqnarray}
  \label{eq:62}
  B_{\alpha\, \gamma|i} =   B_{\alpha \, \gamma} (\a) +  \hat 
B_{\alpha\, \gamma|i},
\end{eqnarray}
where $\hat B_{\alpha \, \gamma|i}$ does not depend on $\a$.
Now $B_{\alpha \, \gamma}(\a)$ cancels out in
(\ref{eq:63}), and (\ref{eq:63}) reduces to (\ref{eq:73}), where
[cf. (\ref{bau})]
\begin{eqnarray}
\label{kara}
&&  \mu_{\sigma>1} = \frac{\hat{{\Omega}}_{\sigma\sigma'}
  \hat{{\Omega}}_{\sigma' ...} \hat{{\Omega}}_{...\sigma''} 
  \hat{{\Omega}}_{\sigma''
    \sigma_0}}
{\hat{{\Omega}}_{\sigma'\sigma}
  \hat{{\Omega}}_{...\sigma' } \hat{{\Omega}}_{\sigma'' ...}
  \hat{{\Omega}}_{\sigma_0
    \sigma'' }},\qquad \mu_1=1,\\
&& \hat\Omega_{\alpha\gamma}=\frac{1}{\hat Z_\gamma[\beta]}{\sum}_i \, e^{\hat
  B_{\alpha\gamma|i}
+\frac{\beta_s}{2}(\hat E_{i\gamma} - \hat E_{i\alpha})-\beta \hat
E_{i\gamma}}. 
\label{lokk}
\end{eqnarray}
Hence the free energy (\ref{eq:42}) applies with $\mu_\alpha$ given by
(\ref{kara}). For (\ref{eq:16}), this free energy differs from
(\ref{eq:42}) due to a choice of the constant ${\cal C}$ in
(\ref{arbitrary}).

Note that the tree topology supports the detailed balance:
$\bar\Omega_{\gamma\alpha}\p_\alpha=\bar\Omega_{\alpha\gamma}\p_\gamma$.
This relates with the Kolmogorov's criterion for the detailed balance:
for all loops of the connection network the product of the transition
probabilities $\bar\Omega_{\gamma\alpha}$ calculated in the clock-wise
direction should be equal to the product in the anti-clock-wise
direction \cite{zia}. For a tree-like network there are no loops,
hence the criterion holds.

For general rates already one loop (i.e., a three-level system) may
suffice for invalidating the existence free energy. But we expect that
loops will be less relevant for multi-dimensional models. Indeed, let
us return to $\p_\alpha\propto \mu_\alpha e^{-\beta E_\alpha}$,
$\p_{i\alpha}\propto e^{-\beta\hat E_{i\alpha}}$, but instead of
(\ref{eq:11}) we assume that only one externally driven parameter (out
of two) pertains to the slow variable: 
\begin{eqnarray}
  \label{eq:6}
E_{i\alpha}(\a)=E_{\alpha}(a_1)+\hat E_{i\alpha}(a_2).   
\end{eqnarray}
Eq.~(\ref{eq:6}) is the minimal situation, where the work-extraction
via a slow, cyclic process is possible.

We get for the
rotor of the work $\w$ [cf.~(\ref{eq:14}) and (\ref{bau})]:
\begin{eqnarray}
  \partial_{a_2}w_{a_1}-  \partial_{a_1}w_{a_2} = \langle
\,(\partial_{a_1}E\,)\,\partial_{a_2}[\,\ln\mu -T\beta_s\ln\hat
Z\,]\,\rangle\nonumber\\
- \langle
\,\partial_{a_1}E\rangle\,\langle \,\partial_{a_2}[\,\ln\mu -T\beta_s\ln\hat
Z\,]\,\rangle, 
  \label{eq:19}
\end{eqnarray}
where $\langle X\rangle\equiv\sum_\alpha\p_\alpha
X_\alpha$. Eq.~(\ref{eq:19}) shows how far is the work from having a
gradient when the partial controllability (\ref{eq:11}) does not
hold. It also determines the amount of work extracted from a cycle in
the $(a_1,a_2)$-space. Now since (\ref{eq:19}) is a correlation, 
the existence of a gradient for $w_{\a}$ can be recovered if
fluctuations are negligible.

% For a particular example of (\ref{eq:63}) consider the simplest tree
% with one root $1$ and three branch-points $2,3,4$.  In accordance to
% (\ref{eq:63}), we directly deduce from (\ref{eq:x5}): $\bar
% p_\alpha=\bar{\Omega}_{\alpha 1}\bar{p}_1/\bar{\Omega}_{1 \alpha}$,
% where $\alpha=2,3,4$.

% The intuition behind the existence of the free energy is that some of
% cycles (closed paths) in the configuration space of the system are
% absent, i.e., we assume that the slow variable lives in a tree-like
% configuration space (this includes one-dimensional situation), while
% other cycles (via the fast degree of freedom) are suppressed due to
% the time-scale separation.

\comment{This system differs from the $N=3$ case of (\ref{eq:8}) by
  the fact its topology is that of a circle, while in (\ref{eq:8}) we
  studied the 1D case with fixed boundaries. To understand the origin
  of this result, note from (\ref{eq:5}--\ref{eq:77}) that generally
  no effective detailed balance holds with respect to $E_\alpha$ and
  $\bar{\Omega}_{\alpha\gamma}$.}

\comment{
A particular, but important, case of (\ref{eq:63}) is the
one-dimensional situation, where a line of $N$ states originates from
the state $1$ (root):

\comment{
\begin{eqnarray}
  \label{eq:8}
  \omega_{\alpha\gamma}=[1-\delta_{\alpha 1}]\delta_{\alpha-1\,\gamma}\,
  \omega_{\alpha \, \alpha-1} +
  [1-\delta_{\alpha A}]\delta_{\alpha+1\,\gamma}\,
  \omega_{\alpha \, \alpha+1},
  \qquad   \alpha=1,...,A, 
\end{eqnarray}
where $\delta_{\alpha\gamma}=1$ if $\alpha=\gamma$, and 
$\delta_{\alpha\gamma}=0$ if $\alpha\not=\gamma$. }

\begin{eqnarray}
  \label{eq:9}
  \bar{p}_\alpha
=\prod_{\gamma=1}^{\alpha-1}\frac{\bar{\Omega}_{\gamma+1\,
    \gamma}  }{\bar{\Omega}_{\gamma\,\gamma+1}}~ \bar p_1.
\end{eqnarray}
The stationary probabilities (\ref{eq:9}) can be written as in 
(\ref{eq:73}) with 
\begin{eqnarray}
  \label{eq:64}
 \mu_\alpha = {\hat Z_\alpha[\beta]}\prod_{\gamma=1}^{\alpha-1}
\frac{\sum_i e^{ \hat B_{\gamma +1 \,
      \gamma|i}+\frac{\beta_s}{2}(\hat E_{i\, \gamma}-\hat E_{i\, \gamma
      +1})-\beta \hat E_{i\gamma}   }}
{\sum_i e^{ \hat B_{\gamma +1 \,
      \gamma|i}-\frac{\beta_s}{2}(\hat E_{i\, \gamma}-\hat E_{i\, \gamma
      +1})-\beta \hat E_{i\,\gamma+1}   }}.
\end{eqnarray}
where $Z_\alpha[\beta]$ is given by (\ref{eq:7}), and $\Z$ is defined
from normalization. 

Recall that for the Kawasaki rates $ B_{\gamma +1 \, \gamma|i}=0$, 1D
master-equation (\ref{eq:x5}) relates to the discrete space
Fokker-Planck equation \cite{agmon_h}.
}

\comment{
\textit{\textbf{ Metastable states}} emerge when the stochastic system
is bound to a part ${\cal M}$ of its configuration set due to kinetic
reasons, i.e., the transition probabilities to the unavailable part
vanish. Once the system spends a long time in ${\cal M}$, it relaxes
to the stationary probabilities restricted to ${\cal M}$; see
\cite{leyvraz} for a recent discussion of equilibrium metastability.
For external processes that slower than this relaxation time, but
faster than the life-time of the metastable state, the work does have
a potential, i.e., the free energy $F_{\cal M}=-\frac{1}{\beta}\ln
\left[ \sum_{k\in {\cal M}}\,e^{-\beta E_k} \right]$ of the metastable
state.

Whenever these transition probabilities become sizable (e.g., due to a
catalyst), a relaxation process starts due to which the system
explores its full configuration set.  The free energy of the stable
state is $F=-\frac{1}{\beta}\ln \left[ {\sum}_{k}\,e^{-\beta
    E_k}\right]$, where the summation over $k$ is taken over all
states.  The fundamental feature of the equilibrium free energy is
that $F\leq F_{\cal M}$. To give $F\leq F_{\cal M}$ an operational
meaning, we need to have a {\it reference state} that is reachable
both from the metastable state and the stable state via a slow
variation of parameter. We discuss this aspect below.

One can try to give a more general meaning to the decrease of the free
energy by considering the well-known H-function $\sum_k p_k
(t)[E_k+T\ln p_k(t)]$ that (at least for the Markov dynamics and
related situations) does decrease monotonously in time during the
relaxation, $p_k(t)\to e^{-\beta E_k}/Z$, to the (unique) equilibrium
state from any initial state (H-theorem) \cite{risken}. We refrain
from doing so, because despite of its suggestive form, it is clear
that the H-function does not generally has the physical meaning of the
potential for work. Indeed, closely related versions of the H-theorem
may be violated for realistic relaxation processes, but this fact is
not taken as an evidence for an anti-thermodynamic arrow of time
\cite{jaynes}.

For the non-equilibrium free energy (\ref{eq:42}), we assume that a
part of the fast configuration set is not available due to
$\rho_{kj|\alpha}\to 0$ for certain $j\in {\cal M}$, $k\not \in {\cal
  M}$, and all $\alpha$; see fig.~1 for an example. Here ${\cal M}$
defines the metastable state; it is a sub-set of 
$1,...,n$ for the fast variable.  We assume that out of ${\cal
  M}$ are much slower than the slow variable.

Using as example the activation rates (\ref{eq:16},
\ref{au}, \ref{eq:42}), we get for the metastable free energy
$\F_{\cal M}$ 
\begin{align}
  \label{bora}
  \F_{{\cal M}}&=-\frac{1}{\beta_s}\ln\left[
    {\sum}_{\alpha}  \mu_{\alpha|{\cal M}}\, e^{-\beta_s
      E_\alpha} \right],\\
  \mu_{\alpha|{\cal M}}&=   {{\sum}_{i\in {\cal M}} e^{-\beta
      \hat  E_{i\alpha}}}\left/  { {\sum}_{i\in {\cal M}} e^{-(\beta-\beta_s)
      \hat  E_{i\alpha}}} \right.
\end{align}
$\F_{{\cal M}}$ is to be compared with the free energy $\F$ of the
stable state; see (\ref{eq:42}). 

Let us now find a sufficient condition for $\F\geq \F_{\cal M}$.  If
for all $\alpha$, and any $ k\not\in {\cal M}$ and $i\in {\cal M}$:
\begin{eqnarray}
  \label{eq:5}
\hat  E_{k\alpha}\geq  \hat E_{i \alpha},~~~ {\rm then}~~~
\mu_{\alpha} \leq \mu_{\alpha|{\cal M}} ~~~{\rm and}~~~
\F_{{\cal M}}\leq \F,
\end{eqnarray}
i.e., the free energy is larger in the stable state. (If $\hat
E_{k\alpha}\leq \hat E_{i \alpha}$, then $\F_{{\cal M}_f}\geq
\F_{{\cal M}_f+{\cal \bar M}_f}$.)

We now construct an example of a reference state and show that|given
constraint (\ref{eq:11})|$\F_{{\cal M}}\leq \F$ is beneficial with
respect of work. Let in $\hat E_{k1}> \hat E_{i1}$ hence
$\mu_1<\mu_{1|{\cal M}}$, but $\hat E_{k\, \alpha}= \hat
E_{i\,\alpha}$ and $\mu_\alpha=\mu_{\alpha|{\cal M}}$ for $\alpha\not
=1$. Thus taking $E_\alpha\to\infty$ we reach a state with the free
energy $\F_{{\cal R}}=-\frac{1}{\beta_s}\ln\left[
  {\sum}_{\alpha\not=1} \mu_{\alpha|{\cal M}}\, e^{-\beta_s E_\alpha}
\right]$ that can be reached both from the stable state and from the
metastable state, but reaching it from the stable state requires less
work, since $\F_{{\cal M}}\leq \F$.

We do not focus on metastability with respect to the slow variable,
since it decreases in the stable state for the same as the equilibrium
free energy does. This relates to the fact that $\sum_\alpha
p_\alpha(t)[E_\alpha+T\ln p_\alpha(t)]$ is an H-function for the slow
dynamics (\ref{eq:x5}), while in the stationary state it equals $\F$;
see (\ref{eq:42}).

}

\textit{\textbf{ Quasi-continuous limit for the slow variable.}}
Recall that the index $\gamma=1,...,N$ numbers the states of the slow
variable. We now consider the case, where these states are arranged
over a segment of a line (1D situation), so that only transition from
one neighbour state to another are allowed. This is an example of
birth-death processes that have many applications
\cite{kampen_book}. If the segment is finite and the states are
homogeneously and densely located in it, then one can pass to the
continuous limit, where instead of a discrete index $\gamma$ we shall
have a continuous variable $x$ \cite{agmon_h,risken}. The continuous
limit is achieved by
\begin{gather}
  \label{eq:12}
  E_{i\,\gamma}\to E_i(x), \quad
  E_{i\,\gamma +1}\to E_i(x)+\hat\epsilon E'_i(x), \\
  B_{\gamma+1\, \gamma|i}\to  B_i(x),
\end{gather}
where $x$ is a continuous parameter, $\hat\epsilon$ is a small
parameter (the distance between the states $\gamma$ and $\gamma+1$ on
the segment), and where $A'(x)\equiv {\d A(x)}/{\d x}$. We now expand
in (\ref{eq:10}) over a small $\hat \epsilon$
\begin{eqnarray}
  \label{bu}
    \omega_{\gamma +1 \, \gamma|i}=\exp\left[B_{\gamma +1 \,
      \gamma|i}\right] (1+{\beta_s}
(E_{i\, \gamma}-E_{i\, \gamma +1})/2),
\end{eqnarray}
and get in (\ref{eq:1}) (see \cite{risken,agmon_h} for similar
derivations): 
\begin{gather}
  \label{eq:26}
    {\sum}_{\gamma}[\omega_{\alpha\gamma|i}\,
  p_{i\gamma} -\omega_{\gamma\alpha|i}\, p_{i\alpha}  ]
=\beta_s[\zeta_{i\,\alpha+1}
  -\zeta_{i\,\alpha}]~~~~~~~~~~~~~~~\nonumber\\
\label{u1}
+\xi_{i\,\alpha+1}
  -\xi_{i\,\alpha}, ~~~~ \xi_{i\,\alpha+1}\equiv
  e^{B_{\alpha+1\,\alpha|i}}\,(p_{i\,\alpha+1}-p_{i\,\alpha}), \\
\label{u2}
\zeta_{i\,\alpha+1}\equiv e^{B_{\alpha\,\alpha+1|i}}\,
    (E_{i\, \alpha+1}-E_{i\,\alpha})\,
    (p_{i\,\alpha}+p_{i\,\alpha+1})/{2}.\nonumber
\end{gather}
Taking in (\ref{u1}) the continuum limit, we get from
(\ref{eq:1}):
\begin{gather}
  \label{eq:29}
  \dot{p}_{i}(x,t)={\sum}_j [\rho_{ij|x}\,p_{j}(x)
  -\rho_{ji|x}\,p_{i}(x,t)
  ]+\epsilon\,\hat\epsilon^2\partial_xJ_i(x,t), \\
  J_i(x,t)\equiv \beta_s \,e^{B_i(x)}\, p_i(x,t)\, E'_{i}(x)
  +e^{B_i(x)}\,p'_i(x,t),
  \label{eq:290}
\end{gather}
where $p_i(x)$ is the joint probability of $i$ and $x$: $\sum_i\int\d
x\, p_i(x,t)=1$, and $J_i(x,t)$ is the probability current related to
the slow variable.

Eq.~(\ref{eq:29}) shows that as compared to (\ref{eq:1}) the slow-fast
limit is facilitated due to the additional small factor $\hat
\epsilon^2$. This is confirmed by fig.~1 which shows that the
slow-fast limit improves when the number $N$ of states of the slow
variable is large. Thus the above continuous limit is a way to get
the time-scale separation naturally.

We treat (\ref{eq:29}) wih the same time-scale separation argument
(\ref{eq:13}--\ref{eq:x5}) with minor modifications for the continuous
case.  The stationary conditional probability is still $\bar
p_{i|x}\propto e^{-\beta E_i(x)}$. For the stationary probability
density $\bar p(x)$|which holds the zero-current condition $\sum_i
J_i(x)=0$|we obtain from (\ref{eq:29}, \ref{eq:290}):
\begin{align}
& \bar{p}(x)\propto e^{-\beta_s E(x)}\mu(x), \qquad 
\mu(x)= e^{h(x)}
{\sum}_i\, e^{-\beta\hat E_i(x)}, \nonumber \\
& h'(x)=(\beta-\beta_s) 
\frac{\sum_i \hat E_i'(x) e^{\hat B_i(x)-\beta \hat E_i(x) }} {\sum_i e^{
    \hat B_i(x)-\beta \hat E_i(x)} },
  \label{eq:55}
\end{align}
where by analogy to (\ref{eq:11}, \ref{eq:62}):
$E_i(x;\a)=E(x;\a)+\hat E_i(x)$, $B_i(x;\a)=B(x;\a)+\hat B_i(x)$.
The free energy is given as in (\ref{eq:42}) by
$\F=-T_s\ln\int\d x\, \mu(x)\, e^{-\beta_s E(x)}$. Note that it goes
to the equilibrium expression for $T_s=T$. 

% Now we can obtain further scenarios of the effect of free energy
% increase due to going from aa metastable to a stable state, e.g., for
% $\hat B'_i=0$, (\ref{eq:55}) integrates directly and predicts the
% effect whenever $\beta$ is sufficiently larger than $\beta_s$, and
% $B_{k\not\in{\cal M}}\geq B_{i\in{\cal M}}$. One can also study more
% scenarios for $\partial_{T_s} \F|_{T}\geq 0$ and $\partial_{T}
% \F|_{T_s}\geq 0$ that was so far obtained only for the activation
% rate; see (\ref{eq:77}, \ref{eq:78}).

\comment{
The choice (\ref{eq:16}) leads in (\ref{eq:15}) to
\begin{eqnarray}
  \label{eq:17}
  B_i(x)=\beta_s E_i(x).
\end{eqnarray}
Hence we get from (\ref{eq:55}, \ref{eq:16}, \ref{eq:17})
\begin{eqnarray}
  \label{eq:18}
  f(x)=\ln\left[\sum_i e^{(\beta_s-\beta)E_i(x) } \right]
  - \ln\left[\sum_i e^{-\beta E_i(x) } \right],
\end{eqnarray}
which is the difference of two free energies at different inverse
temperatures. }

% \textit{\textbf{ Non-equilibrium free-energy with equilibrium
%     features.}}  While we focussed on anomalous features of the
% non-equilibrium free energy, it is useful to recall one particular
% situation, where its features are in in accord to equilibrium
% expectations.

\comment{
\textit{\textbf{ Metastable states}} emerge when the stochastic system
is bound to a part ${\cal M}$ of its configuration set due to kinetic
reasons, i.e., the transition probabilities to the unavailable part
vanish ${\cal S}$. The system spends a long time in ${\cal M}$ and
gets into the stationary state restricted to ${\cal M}$. But once
after a long observation time the system spontaneously moves from
${\cal M}$ to ${\cal S}+{\cal M}$, the contribution of ${\cal M}$ into
the stable state of the system is neglegible, because all the energies
in ${\cal M}$ are much larger than those in ${\cal S}$; see
\cite{leyvraz} for a recent discussion.

The equilibrium free energy decreases when going from a metastable
state to a stable state: $-T\ln [\sum_{k\in {\cal M}}e^{-\beta E_k}]>
-T\ln [\sum_{k\in {\cal S}+{\cal M}}e^{-\beta E_k}]$.

For the non-equilibrium situation (\ref{eq:55}) we can have
metastability with respect to the fast variable, where (for all
$\alpha$'s) the set of energies $\{E_{k \alpha}\}_{k\in {\cal M}}$
spontaneously enlarges to $\{E_{i\alpha}\}_{i\in {\cal M}+{\cal
    S}}$. Time-scale separation holds both within the metastable state
(since the slow transitions of the fast variable are blocked), and
within the stable state, because ${\cal M}$ is irrelevant for it.  But
the non-equilibrium free energy $\F$ given by (\ref{eq:55}) can be
larger in the stable state; e.g., take $\partial_x \hat B_i(x)=0$ and
$\beta\gg \beta_s$ in (\ref{eq:55}). The situation here is similar to
the behavior of $\F$ with respect to $T$ [cf.~(\ref{eq:78})], because
a bigger $T$ also means a larger configuration set of the fast
variable.

\comment{
For external processes that slower than this relaxation time, but
faster than the life-time of the metastable state, the work does have
a potential, i.e., the free energy $F_{\cal M}=-\frac{1}{\beta}\ln
\left[ \sum_{k\in {\cal M}}\,e^{-\beta E_k} \right]$ of the metastable
state.

Whenever these transition probabilities become sizable (e.g., due to a
catalyst), a relaxation process starts due to which the system
explores its full configuration set.  The free energy of the stable
state is $F=-\frac{1}{\beta}\ln \left[ {\sum}_{k}\,e^{-\beta
    E_k}\right]$, where the summation over $k$ is taken over all
states.  The fundamental feature of the equilibrium free energy is
that $F\leq F_{\cal M}$. To give $F\leq F_{\cal M}$ an operational
meaning, we need to have a {\it reference state} that is reachable
both from the metastable state and the stable state via a slow
variation of parameter. We discuss this aspect below.}
}

\textit{\textbf{ Non-equilibrium free-energy with equilibrium
    features.}} A particular case of (\ref{eq:55}) is when $B_{i}(x)$
does not depend on $i$, i.e., $\hat B_{i}(x)=0$. Now the
non-equilibrium free energy reads \cite{an}:
\begin{eqnarray}
  \label{eq:69}
\F^\circ=-T_s\ln\left[
\int \d x\, (\, {\sum}_i\, e^{-\beta E_i(x)}\,)^{{\beta_s}/{\beta}}
\right].
\end{eqnarray}
The following features of $\F^\circ$ are deduced directly from
(\ref{eq:69}) \cite{an}. They all are very similar to those of the
equilibrium free energy.

-- $\F^\circ$ is the potential for the work {\it without} restriction
(\ref{eq:11}), i.e., now $\a$ can also enter $\hat E_i(x)$:
\begin{eqnarray}
  \label{eq:111}
  E_{i}(x;\a)=E(x;\a)+\hat E_{i}(x;\a).
\end{eqnarray}

-- For the temperature-derivatives of we get
\begin{eqnarray}
  \label{eq:71}
  \partial_T \F^\circ|_{T_s}= 
  -S,  \qquad
  \partial_{T_s}   \F^\circ |_{T}= 
-S_{ s}, 
\end{eqnarray}
where $S\equiv -\int\d x\, {\sum}_k \p_k(x) \ln \p_{k|x}$ and $S_{
  s}\equiv -\int\d x\, \p(x) \ln \p(x)$ are, respectively, the
conditional entropy of the fast variable, and the marginal entropy of
the slow variable. Hence the anti-thermodynamic cooling effect noted in
(\ref{wot}) is impossible here.

% -- $\F^\circ$ is larger for metastable states (compared to stable
% states), for the same mechanism as in the equilibrium case.

Note the analogy between (\ref{eq:71}) and the equilibrium formula
$\partial_T F_{\rm eq}=-S_{\rm eq}$. It leads us to $\F^\circ = U-T_s
S_{ s}-TS$, where $U= \int\d x\, {\sum}_k \p_k(x)E_k(x)$ is the
average (overall) energy. This expression for $\F^\circ$ already
appeared in the physics of Brownian motion \cite{landauer,an} and
glasses \cite{theo}.

Eq.~(\ref{eq:69}) has an intuitively appealing meaning
\cite{landauer,an,sg1,sg2,sg3,karen,figur,ilg}, since it implies that
the free energy $-T\sum_i e^{-\beta E_i(x)}$ of the fast variable
(evaluated at a fixed value of the slow variable) serves as an
effective potential $U_{\rm eff}(x)$ for the (Gibbs) distribution of
the slow variable: $\p(x)\propto e^{-\beta_s U_{\rm eff}(x)}$. And
then $\F^\circ$ is the free energy related to that effective Gibbs
distribution: $\F^\circ=-T_s\ln\int\d x\,e^{-\beta_s U_{\rm
    eff}(x)}$. While this heuristic explanation is frequently applied
in statistical physics, we should keep in mind from the above
derivation that $\F^\circ$ exists {\it due} to the continuous limit,
e.g., assuming $\hat B_{\alpha\gamma}=0$ in the discrete case does not
recover the above equilibrium features (for $T\not=T_s$).

\textit{\textbf{ Summary.}} We studied a class of non-equilibrium
stationary states generated by two thermal baths at different
temperatures. Three conditions were assumed: 

{\it (i)} Variables of the system have different characteristic times.

{\it (ii)} External fields act only on the slow variable.

{\it (iii)} The transition rates of the slow variable are such that
(under time-scale separation) it holds an effective detailed balance.
Examples of the effective detailed balance are the activation
transition rate for an arbitrary topology of connections, and
arbitrary rates for a tree-like topology.  

Conditions {\it (iii)} are similar to those governing the no-pumping
theorem \cite{p1,p2,p3,p4,p5}. It considers a stochastic system
coupled to a single thermal bath|but driven by an oscillating external
field|and studies conditions under which the time-averaged probability
currents vanish.  The no-pumping theorem has extensions beyond of the
activation rates and loop-less networks \cite{narek}.

Under conditions {\it (i)--(iii)} there exists a non-equilibrium free
energy. It describes quasi-stationary isothermal processes, and is
defined via the potential of the work done via a slow variation of
external parameters. As compared to its equilibrium counterpart, the
non-equilibrium free-energy can be an increasing function of the bath
temperature. As a physical demonstration of this feature we studied
the free-energy cost of the isothermal cooling (i.e., entropy decrease
via slow external fields): this cost can decrease if the cooling
starts from a higher temperature. It is interesting to compare this
finding with the Mpemba phenomenon \cite{kell,linden}: given
two samples of water that are identical except of their initial
temperature, the initially hotter samples cools (and freezes) quicker
than the colder one. This can have relatively straightforward physical
explanations, e.g., some part of the hotter water can evaporate thus
decreasing its amount and making its cooling easier. The Mpemba phenomenon
disappears once such scenarios are ruled out \cite{linden}. To compare
our finding with the Mpemba effect, we first of all note that our
cooling is not spontaneous. We focus on an (isothermal) entropy
reduction of the system by means of external fields. (Still our
cooling process is slow and it leaves the system in its locally
stationary state.) Hence we study not the speed of cooling, but its
work cost, as determined by the free-energy. The effect we found is
impossible in equilibrium, because the equilibrium free energy is a
decreasing function of the temperature. However, we show that this
effect survives close to equilibrium. Thus, the second difference
with the Mpemba phenomenon is that our effect is out of
equilibrium, as it relates to two different temperatures.

The existence of the free energy is facilitated in the
(one-dimensional) continuous limit of the slow variable. Now
time-scale separation comes out naturally, and there is a situation,
where a non-equilibrium free energy exists for all driven parameters,
i.e., assumption {\it (ii)} can be relaxed.

The free energy allows to involve ideas of equilibrium
thermal physics for understanding non-equilibrium
stationary states; other approaches that aim at a
thermodynamical description of such states are studied
in \cite{keizer,ben,blythe,paquette,robinson,komatsu,ge,bouchet}.

% This result generalizes our previous finding of a non-equilibrium free
% energy in such two-temperature systems \cite{an}. The generalization
% is substantial, because the free energy found in \cite{an} had almost
% all features of the equilibrium free energy.

\textit{\textbf{ Acknowledgements.}}
This work is partially supported by COST MP1209 and by the ICTP
through the OEA-AC-100.

%%%%%%%%%%%%%%%%%%
%%%%%%%%%%%%%%%%%%

%%%%%%%%%%%%%%%%%%
%%%%%%%%%%%%%%%%%%
\comment{

}
%Thermodynamics at nonequilibrium steady states

% Heat, work, and the thermodynamic temperature at nonequilibrium
% steady states 

% –565 The freezing of hot and cold water. 

%–885 Supercooling and the mpemba effect: When hot water freezes quicker
% than cold.

%–522 The mpemba effect: When can hot water freeze faster than cold? 

%%%%%%%%%%%%%%%%%%
%%%%%%%%%%%%%%%%%%

\comment{
\clearpage 

\vspace{2.5cm}

{\bf\large Response to Referee A}

\vspace{0.5cm}

We are grateful to Referee A for the encouraging report. The remarks
by Referee A are reproduced in italics and replied below.

1. {\it I have witnessed that under certain circumstances, hot water
  pipes freeze earlier than adjacent cold water pipes. Can this be
  taken as motivation for the work, and is the case covered under the
  first section of page 4?}

Referee A is presumably referring to the so called Mpemba phenomenon
\cite{kell1,linden1}: hot water cools (and freezes) quicker than a
colder one. (It is assumed that both samples of water are identical
except of their initial temperature.) The phenomenon was observed
independently under various circumstances; apparently, its
existence was known already to Aristotle \cite{linden1}. It can have
relatively straightforward physical explanations, e.g., some part of
the hotter water can evaporate thus decreasing its amount and making
its cooling easier. Another explanation is that a fast cooling can
force the water out of local equilibrium \cite{lu1}.

Once such straightforward scenarios are ruled out experimentally, the
Mpemba phenomenon disappears \cite{linden1} (nobody was so far able to
replicate the original data by Mpemba and Osborne). It is still
possible that the phenomenon survives due to a different chemical
composition of a heated water (which may lead to a different
concentration of dissolved gases) \cite{katz1}. At any rate, the
phenomenon is not observed once one ensures the identical chemical
composition of both samples, e.g., by boiling them first and only then
setting them to different initial temperatures \cite{linden1}.

Let us now discuss how or result differs from the above Mpemba
phenomenon (without predetermining its causes).

First of all, our cooling is not spontaneous. We focus on an
(isothermal) entropy reduction of the system by means of external
fields. (Still our cooling process is slow and it leaves the system in
its locally stationary state.) Hence we study not the speed of
cooling, but its work cost, as determined by the free-energy. We find
that this cost is smaller, if one starts to cool at a higher
temperature. Such an effect is impossible in equilibrium, because the
equilibrium free energy is a decreasing function of the
temperature. However, we show that this effect survives close to
equilibrium.  Hence the second major difference with the Mpemba
phenomenon is that our effect is explicitly out of equilibrium, as it
relates to two different temperatures.

To our understanding, the Mpemba phenomenon is sufficiently different
from our result, but it is still very interesting to discuss this
phenomenon and to compare it with our finding. This is now done in the
last section of the revised version.

\vspace{0.3cm}

2. {\it Finally, a few very small details:

line before Eq (8): add ", so that"

lowest line of left column page 2 and also further on: "is got" $\to$
"is gotten"

Alinea before (16): "Now ..."  Can you add a few more words to make it
more clear?  Next line: "holds" $\to$ "implies"; also 2 lines before
(40).

Around Eq (29): replace index "1" by $\sigma_1$ or $\sigma_0$.

First line of last alinea: skip the second "the"}

These suggestions were followed in the revised version. In particular,
we employed $\sigma_0$, and also explained in more detail the alinea
before (former) eq. (16).

\vspace{0.5cm}

{\bf\large Response to Referee B}

\vspace{0.5cm}

We are grateful to Referee B for thex constructive report that helped
us to revise our manuscript. Below we respond in detail to all comment
and criticisms that are reproduced in italics.

\vspace{0.3cm}

1. {\it The authors seem unaware that the approximation of a slow-fast
  system tracing out fast degrees of freedom is a widely studied
  problem from a rigorous point of view. They can consult the recent
  book by Pavliotis and Stuart on the subject: ‘Multiscale methods:
  averaging and homogenization’. In this sense, it is unclear why the
  authors present Fig. 1. The fact that solutions to eq. (1) are well
  approximated in the limit of $\epsilon\ll 1$ by the solution of the
  corresponding slow system, is rather obvious and even rigorously
  known in my opinion (if caveats are present, the authors should
  explain them). }

Thank you for recommending this book to us. We were not aware about
it, but now we have studied its relevant parts and did benefit from
it. It is true that Pavliotis and Stuart present rigorous results on
the averaging of the fast-slow Markov chains that are highly relevant
for us. Hence the book is cited and discussed in the present version
of the manuscript. In particular, it is mentioned that the approach of
the book is a rigorous one. One aspect that was not explicitly
stressed in the book (and which has to do with the next question by
Referee B) is that in the stationary situation the full stationary
probability of the Markov system can be found via the averaging
approach. However, this task can be solved via basically the same
perturbative approach, as we now explicily demonstrate in the revised
version; please eqs. (5--9).

The purpose of presenting Figs.~1 is as follows. Both figures study a
concrete (but to our experience rather generic!) numerical
examples. They show that zeroth order result of the time-scale
separation approach can apply beyond the small $\epsilon$
situation. More details on this are presented below, when responding
to the next question by Referee B.

\vspace{0.3cm}

2. {\it It is known (see the above cited book) that solutions of the
  effective slow system approximate solutions of the original one in
  average (averaging) and up to capturing Gaussian fluctuations if
  homogenization methods are used. Now, the authors seem to assume
  that the approach allows to estimate the full probability
  distribution function of the process, see eq. (10) to be valid. This
  is generically not true.  The authors should make clear that they
  are doing such an approximation at least. It would also be welcome
  an estention of their study to understand what is the error
  committed in such an approximation. }

We completely agree with Referee B that generally the result expressed
by eq.(10) does not hold for the slow-fast Markov system. However, we
stated this result only for the stationary distribution. The validity
of this result can be shown via a perturbative approach akin to that
explained by Pavliotis and Stuart. In the present version we carried
out this task and have shown that the error of the approximation is of
order of $\epsilon$, for a small $\epsilon$. Fig.~1 (right sub-figure)
confirms this scaling of the error, but it also shows that for a
larger values of $\epsilon$ the error scales sub-linearly over
$\epsilon$. Since this point is rather beneficial for the validity of
the approximation, and since (to our numerical experience) it is
generic, we decided to illustrate it with Fig.~1.

In the new version, the part devoted to the above points has been
completely re-written to address all the points raised by Referee B. 

\vspace{0.3cm}

3. {\it One of the main assumption of the authors is that the system
  under consideration has a tree-like structure in the slow degrees of
  freedom. The presentation is very abstract and I cannot grasp which
  system, if any, has such a structure. The authors should give
  physical, even if minimal, examples.}

We reworked the presentation of this part and added a figure (Fig.~2)
that should illustrate the simplest examples of the tree-like
structures. One of them considers state of the slow variable that live
on a segment of a line. We hope this aspect is clear by now.

\vspace{0.3cm}

4. {\it In the last part of the paper, the authors pass to the
  continuous limit for the slow degrees of freedom. I really cannot
  grasp what they are trying to do. Indeed, $\gamma$ denotes a state
  of slow variables. In passing to the continuous limit, they seem to
  assume a spatial structure. (they even talk about a ‘lattice
  spacing’ and write things like $\gamma + 1$ which I do not see how
  they make sense). Let me be more precise: if we were considering a
  lattice model (such as ASEP for example) of size $N$, $\gamma$ would
  be a string of 0 and 1, with 0 representing a void site and 1 an
  occupied one. What $\gamma + 1$ would mean here? What would $x$?}

We understand that our previous presentation of the continuous limit
has been patchy. We re-worked it in the new version. We assume that
the states of the slow variable are located on a segment of a line, so
that only transitions between neighbour states are allowed. The index
$\gamma=1,...,N$ numbers the states of the slow variable. This is a
birth-death process, a well-known example of a Markov set-up
\cite{kampen1}. We shall now give an example of it using a
two-particle, two-site situation, where each site can be empty or
occupied. Here we have 4 states; hence $\gamma=1,2,3,4$. These states
read, respectively

$$ (0,0),~~ (0,1),~~ (1,0),~~ (1,1),  $$
where $1$ means that the site is occupied and $0$ means that it is
empty. If we now assume that particles can enter and leave the system
only through the second site, then transitions can happen only from
one neighbour state to another, e.g., there are only transitions:
$ (0,0) \Leftrightarrow (0,1)$, $ (0,1) \Leftrightarrow (1,0)$, 
$ (1,0) \Leftrightarrow (1,1)$. Now if $\gamma=1$ means $ (0,0)$, then 
$\gamma=2$ refers to $(0,1)$ {\it etc}.

This was a simple example. The actual application range of the
birth-death processes is much larger. In the present version we added
a new reference \cite{kampen1}, where many of such applications are
studied.

The meaning of the continuous limit is as follows. We assume that the
segment is finite and the number of states $N$ (of the slow variable)
is large $N\gg 1$. Moreover, the state are located homogeneously,
i.e., one can pass from a discrete index $\gamma$ to a continuous
variable $x$. This transition is well-discussed in the literature
\cite{agmon_h1,risken1}. It produces the Fokker-Planck equation
\cite{risken1}.

\vspace{0.3cm}

5. {\it This might be a minor point if it only result from
  misprints. At [Pag.3, Col.1], the authors state: Cooling amounts to
  an isothermal process, where a(t) slowly changes from a in to a f
  and achieves a lower final entropy [...] I have no idea what the
  authors mean here. For me cooling is a synonimous of lowering the
  temperature, not of lowering the entropy (and a brief google search
  confirmed my impression). }

We agree that the previous version did not properly motivate the
studied definition of cooling. This drawback is corrected in the
revised version. 

It is true that the original meaning of cooling refers to decreasing
the temperature. This requires a system that has well-defined
temperatures both initially and finally (i.e., an equilibrium system),
and then cooling also means that the final entropy (together with the
final temperature) is smaller than the initial one.

This definition was generalized once people got interested by systems
that can be in non-equilibrium states, and/or by systems permanently
coupled to fixed-temperature thermal baths
\cite{nmr_cooling1,ket1,kosloff1,fernan_cooling1,jan1,dvira1,popo1}. In
many practical applications it is desirable to ``localize'' such
systems in their configuration space, i.e., to reduce their
entropy. And since it is normally impossible to change their thermal
baths|or to reduce temperatures of such baths|the entropy reduction is
to be done by means of external fields
\cite{nmr_cooling1,ket1,kosloff1,fernan_cooling1,jan1,dvira1,popo1}.
This is normally more feasible. Please note that in such a process the
bath temperatures stay constant.

Entropy reduction by means of external fields is frequently called
simply cooling
\cite{nmr_cooling1,ket1,kosloff1,fernan_cooling1,jan1,dvira1,popo1},
though one should stress that the cooling is understood in a
generalized sense. One important example of this type occurs in NMR
spectroscopy, where it is highly desirable to increase the spin
polarization, which is the same thing as to reduce the spin entropy
\cite{nmr_cooling1,ket1,fernan_cooling1}. There are connections
between the cooling in the traditional sense and the entropy reduction
by means of external fields \cite{jan1}; please see the response to
the next question.

Thus in the present article, the cooling is defined as entropy
reduction by means of external fields. In the revised version we
introduced new references
\cite{nmr_cooling1,ket1,kosloff1,fernan_cooling1,dvira1,popo1} to
motivate and explain this definition. We also provided an additional
paragraph, where this definition is discussed and motivated.

\vspace{0.3cm}

6. {\it Possibly, one could consider cooling as decreasing the
  internal energy of the system, which I do not see how it should be
  decreasing the entropy.}

The considered set-up can be (under few natural conditions) regarded
as decreasing the internal energy of the system. This relation is not
accidental, it connects to the fact that the fast system is in
conditional equilibrium. In the revised version we added a paragraph
on this point; please see around eq.~(29). 

\vspace{0.3cm}

7. {\it By the way, in other papers already published such as [28],
  the authors seem to give to cooling the natural meaning state
  above. If the authors really mean what they state, they should
  either explain with much more care what they are talking about. If
  not, they should make clear how the following computation in
  eq. (22) supports the claim.}

It is true that in the above reference [28] (which is cited below as
\cite{jan1}), we not only focussed on the cooling as entropy
reduction, but also saw a relation between this definition and
the more standard one, which amounts to temperature reduction.  This
relation is established for equilibrium macroscopic systems that can
be detached from their baths, and for which (equilibrium) thermally
isolated processes are feasible.

These constructions that are not feasible in the context of the
present work, first of all because the studied system is not an
equilibrium one (it couples with two different baths).  Since our
present purpose is to illustrate the fact that the non-equilibrium
free energy can increase with the bath temperature, we decided to
focus on the cooling as entropy reduction via external fields, also
because it has a huge number of applications partially outlined in
\cite{nmr_cooling1,ket1,kosloff1,fernan_cooling1,jan1,dvira1,popo1}.

\vspace{0.3cm}

8. {\it The authors claim to prove that, in nonequilibrium, cooling
  from high temperature $T_h$ of an amount $\Delta T$ might have a
  lower work cost than cooling from a lower one $T_l$. In order to do
  so, they assume that in the parameter change ${\bf a}_i \to {\bf
    a}_f$ , the system changes its temperature by an amount $\Delta
  T$. However, I do not see why this should be true. If not true, such
  an argument invalidate the main result of the paper.}

We consider the bath temperatures $T$ and $T_s$ fixed (isothermal
processes). Our precise statement is as follows: consider the first
set-up, where for a given temperatures $T$ and $T_s$, and a given
(slow) parameter change ${\bf a}_i \to {\bf a}_f$, the entropy is
reduced to small value (and also the internal energy decreases).  This
demands a certain work cost, which calculated via the free energy
difference. Next, consider the second set-up that differs from the
first by the fact that the fixed temperatures are $T+\Delta T$ and
$T_s$, where $\Delta T> 0$. i.e., one of the initial temperatures is
larger. It now appears that the work cost for the second set-up is
smaller. Such an effect is impossible in equilibrium|i.e., for a system
coupled to a single thermal bath|because the equilibrium free energy
(that defines the work cost) is a decreasing function of temperature.

\vspace{0.3cm}

9. {\it The authors should cite the above mentioned book of Pavliotis
  (or any other equally relevant reference) about averaging methods.}

We cite this useful book in the new version. Along with it we cite the
review paper \cite{kampen2} that gives a physical perspective on
slow-fast systems. 

\vspace{0.3cm}

10.{\it Pag.1, Col. 2: Under these conditions the slow (isothermal)
  work done on this system admits a potential, i.e., there exists the
  free energy. This definition is unambiguous, because the work is
  defined for an arbitrary non-equilibrium state. The authors should
  cite a reference for such a statement.}

We now cite \cite{balian1} in this context, and we also explain after
eq.~(15) what do we mean precisely by this statement. We mean to say
that for an arbitrary (not necessarily slow) process described by
master equation (1) the work can be calculated by an analogue of (15),
where $p_{i\alpha}(t)$ is the time-dependent probability. As explained
in \cite{balian1} and also in \cite{lindblad1}, this definition of
work is even more general.

\vspace{0.3cm}

11. {\it Pag.2, Col. 2: Recall that relations between thermodynamic
  and mechanic work have to be fixed in the context of concrete
  applications [29]. What do the authors mean here? They should be
  more precise.}

We added a paragraph that should hopefully clarify the point. It has
to do with the fact that the definition of work (14) has a certain
freedom, which is akin to the gauge-freedom of the potential energy in
mechanics.  In (14) one notes that $\partial_{\a} E_\alpha$ (and hence
$\w$) will change upon adding to $E_\alpha$ a factor $\varphi (\a)$
that does not depend on $\alpha$, but depends on $\a$. Hence the
energies $E_\alpha$ need to be specified properly before calculating
the work.

This point may seem very straightforward (if not trivial), but
forgetting about it led to some confusion in literature. Hence we
decided to mention it.

}

\end{document}